\documentclass[a4paper,11pt]{article}
\usepackage{pos}

\title{``Star coverage'': a simple tool to schedule an observation when FOV rotation matters}
\ShortTitle{``Star coverage'': when FoV rotation matters }

\manuallySeparateAuthors
\author*[a,b]{Simone~Iovenitti}
\author[c]{ for the ASTRI Project}

\affiliation[a]{Università degli Studi di Milano, dip.to di Fisica,\\
Via Giovanni Celoria 16, 20133 Milano, Italy}

\affiliation[b]{INAF--Osservatorio Astronomico di Brera,\\
Via E. Bianchi 46, 23807 Merate, Italy}

\affiliation[c]{\protect\url{astri.inaf.it}}

\emailAdd{simone.iovenitti@inaf.it}

\abstract{
During a tracking mode observation, every telescope with an alt-azimuthal mount shows a rotation in the field of view (FoV) due to the diurnal motion of the Earth. The angular extension of the rotation depends mainly on the time-length of the observation, but also on the telescope's latitude and pointing, because it is determined by the evolution of the parallactic angle of the target,  which is a function of those two parameters.
In many cases, the rotation of the FoV can be exploited to assess some optomechanical properties of the telescope, e.g. the alignment of the optical elements or the motors’ precision during the tracking. As a consequence, it could happen that a proper simulation of the FoV rotation is crucial to program an observation aiming at calibrating the whole system.
We present a tool to simulate the apparent rotation of the FoV, calculating the actual “star coverage” exploitable for scientific goals. Given the FoV and the pointing direction, the software calculates the angular extension of the rotation, considering only the stars observable by the telescope below the magnitude limit. 
This tool will be adopted to schedule the pointing calibration runs of the innovative ASTRI-Horn Cherenkov telescope, developed by INAF for gamma-ray ground-based astronomy, but with the potentiality to produce sky images as an ancillary output, using the so-called Variance method. By exploiting the FoV rotation with the Variance method, the critical assessment of the camera axis can be successfully performed.
}

\FullConference{37$^{\rm{th}}$ International Cosmic Ray Conference (ICRC 2021)\\
		July 12th -- 23rd, 2021\\
		Online -- Berlin, Germany}

\begin{document}
\maketitle

\section{Introduction}
\noindent
Cherenkov telescopes are instruments designed to detect nanosecond light flashes produced in the UV range by the atmosphere (Cherenkov effect) when a gamma-ray photon from the cosmos hits our planet. Because of this peculiar target, the optical scheme and the acquisition process of Cherenkov instruments are usually quite different from those of optical telescopes. In particular, Cherenkov astronomy is generally characterized by a moderate angular resolution (few arcminutes) and a low sensitivity for steady or slow-varying celestial sources \cite{evolution_hillas}. For example, usually the flux from the stellar component of the night sky background is even eliminated using both hardware and software techniques, in order to avoid spurious contributions in the images of Cherenkov light flashes. As a consequence, this process results in telescopes which are actually blind to the star field.
From the point of view of the calibration and assembly verification, the possibility to obtain an image of the stars in the field of view of the telescope would be desirable, as it would be the only chance to see what (and how) the camera is actually observing. For this reason, some Cherenkov telescopes are endowed with pipelines or auxiliary hardware to obtain such information. An example is the Italian project ASTRI (acronym of \textit{Astrofisica con Specchi a Tecnologia Replicante Italiana}), currently aimed at implementing an array of 9 Cherenkov instruments (MINIARRAY), after the success of the prototype telescope ASTRI-Horn, located on Mount Etna in Italy \cite{ASTRI_validation_2019}.
\begin{figure} %
\centering
    \includegraphics[width=0.5\textwidth]{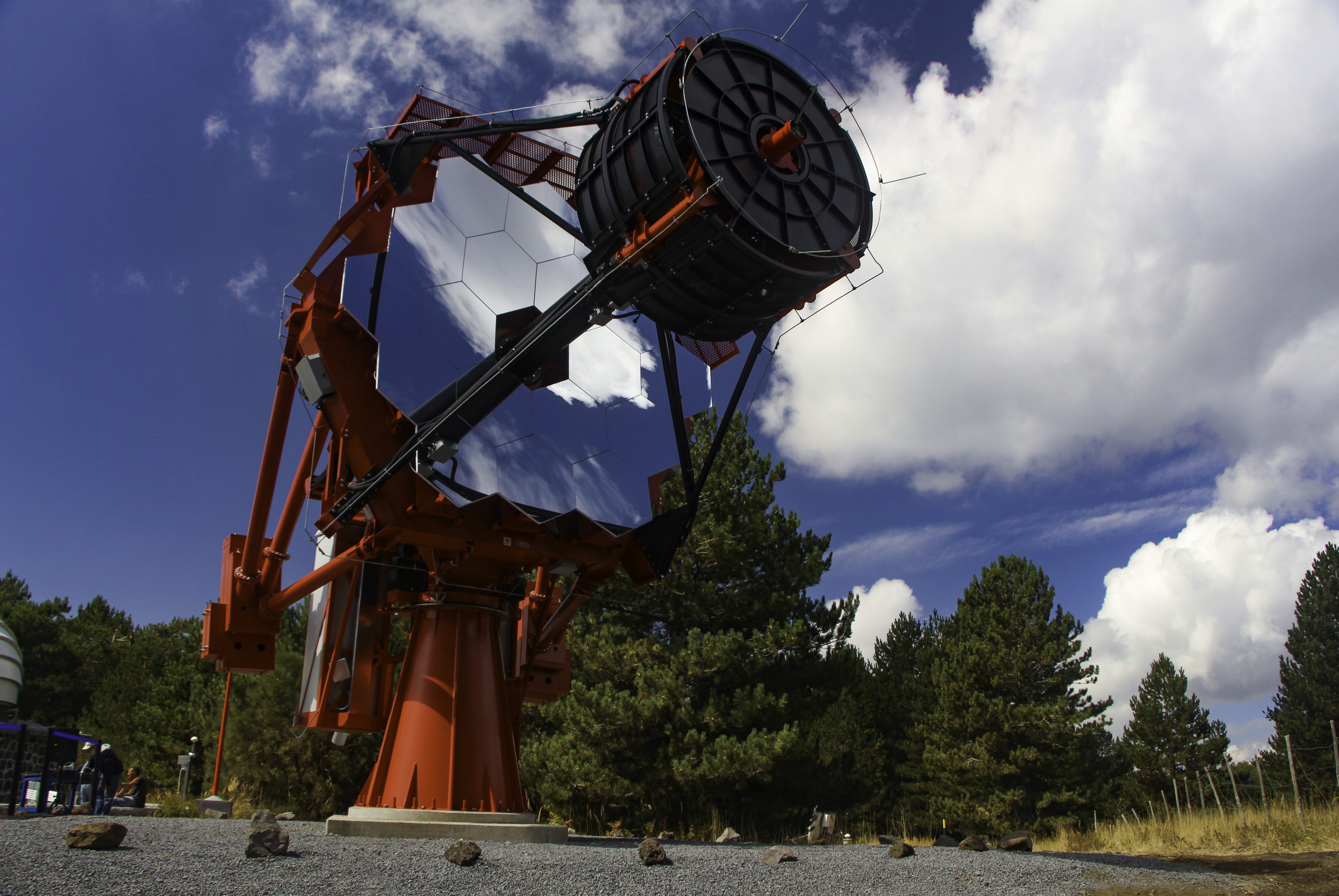}
    \caption{The ASTRI-Horn telescope, located on Mount Etna, in Italy (credit: E. Giro).}
\label{fig_1}
\end{figure}
The novel Cherenkov camera of ASTRI, endowed with miniaturized SiPM sensors (7 mm side), implements a statistical method in its logic board to provide a measure of the photon flux impinging on each pixel as an auxiliary output of the camera. This technique is the so-called “Variance” method: the telescope can observe the Cherenkov flashes, while the Variance produces images of the star field. Thanks to this strategy, ASTRI has the capability to exploit the Variance images to perform a continuous monitoring of several quantities: the pointing direction, the accuracy of tracking, possible anomalies in the gain level of some pixels, the presence of bright objects in the FoV (e.g. satellites) and lots more \cite{Segreto_calibration}. However, because of the modest angular resolution of the Cherenkov camera, especially to measure astrometric quantities (e.g. pointing or tracking) it is necessary to consider long time intervals (few minutes, up to few hours) in order to mitigate the intrinsic modest angular resolution of the Cherenkov camera. With this approach, it is possible to assess the opto-mechanical behavior of the telescope directly from the Cherenkov camera, without any additional hardware, but at the cost of a typical effect of long exposures: the FoV rotation. In the next session we will briefly discuss this phenomenon, pointing out that it is not necessarily negative. In fact, it can be used to optimize telescope performances, especially for Cherenkov instruments. For this reason, in the context of the ASTRI project, we developed a simple software tool to simulate the FoV rotation in detail: “Star Coverage”. The code is open source, and its workflow is described in section 3, together with some practical examples. Finally, in section 4, we outline the usage of Star Coverage in the context of the ASTRI Mini-Array.

\section{The rotation of the FoV and its possible exploitation}
\begin{figure} %
\centering
    \includegraphics[width=0.4\textwidth]{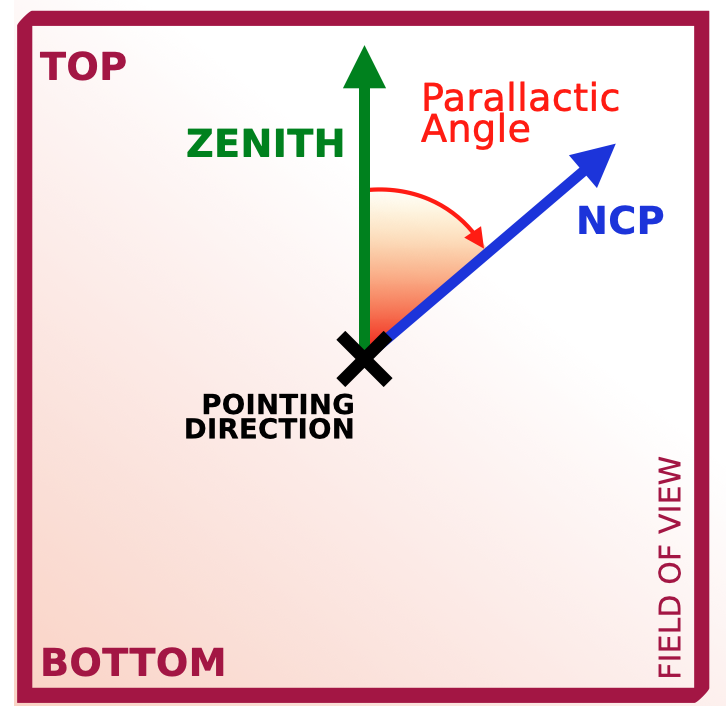}
    \caption{Drawing representing the parallactic angle in the field of view.}
\label{fig_2}
\end{figure}

\noindent
Every telescope with an alt-azimuth mounting system, during a tracking mode observation, presents the effect of FoV rotation: celestial sources apparently rotate around the pointing direction. This effect is due to the orientation of the camera with this particular mount system: the bottom side is parallel to the ground and the top is always oriented towards the zenith. As a consequence, the camera does not rotate according to the movement of the celestial sphere (whose axis is aligned with the North Celestial Pole, NCP) and hence the orientation of the star field with respect to the camera will change in time, resulting in an apparent rotation around the pointing direction (camera center) during tracking observing runs. This effect is completely described by the evolution of the parallactic angle $\eta$, which is the angle between the NCP and the local zenith, with vertex in the pointing direction. Figure~\ref{fig_2} reports a drawing to show the elements involved.

\begin{figure} %
\centering
    \includegraphics[width=\textwidth]{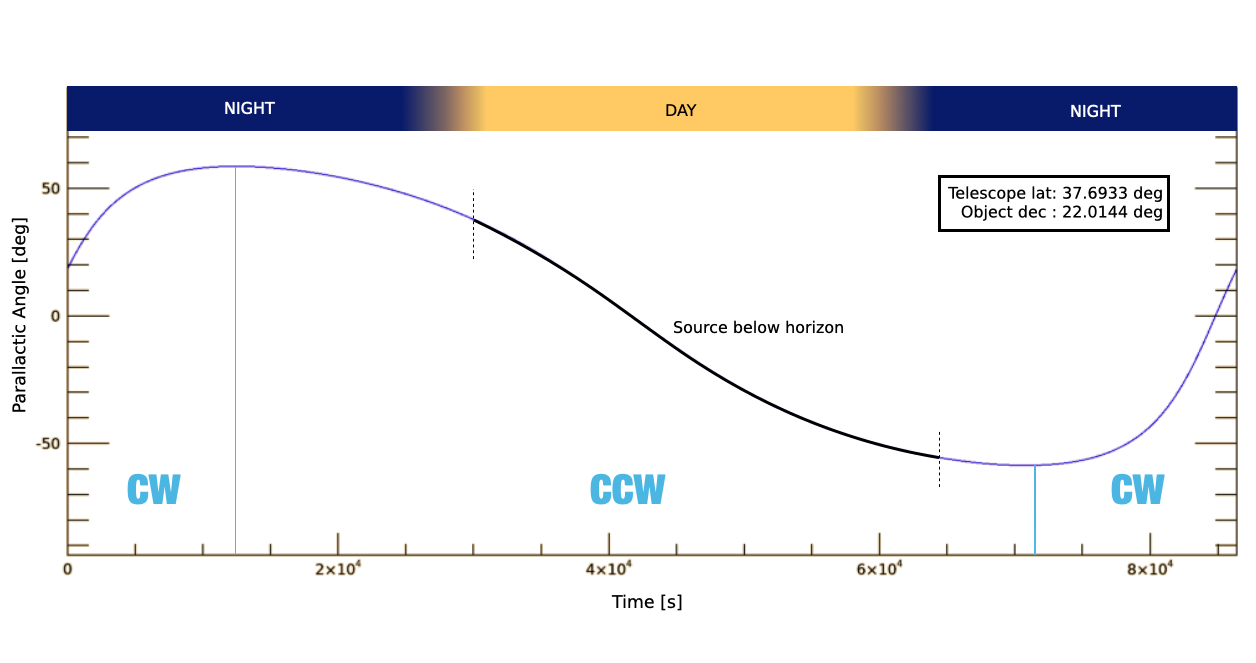}
    \caption{Example of the evolution of the parallactic angle as a function of time for a non-circumpolar source in a virtual acquisition of 24 h. The direction of rotation changes from clockwise (CW) to counterclockwise (CCW) while the source is above the horizon (blue line) during the night.}
\label{fig_3}
\end{figure}

\begin{figure} %
\centering
    \includegraphics[width=\textwidth]{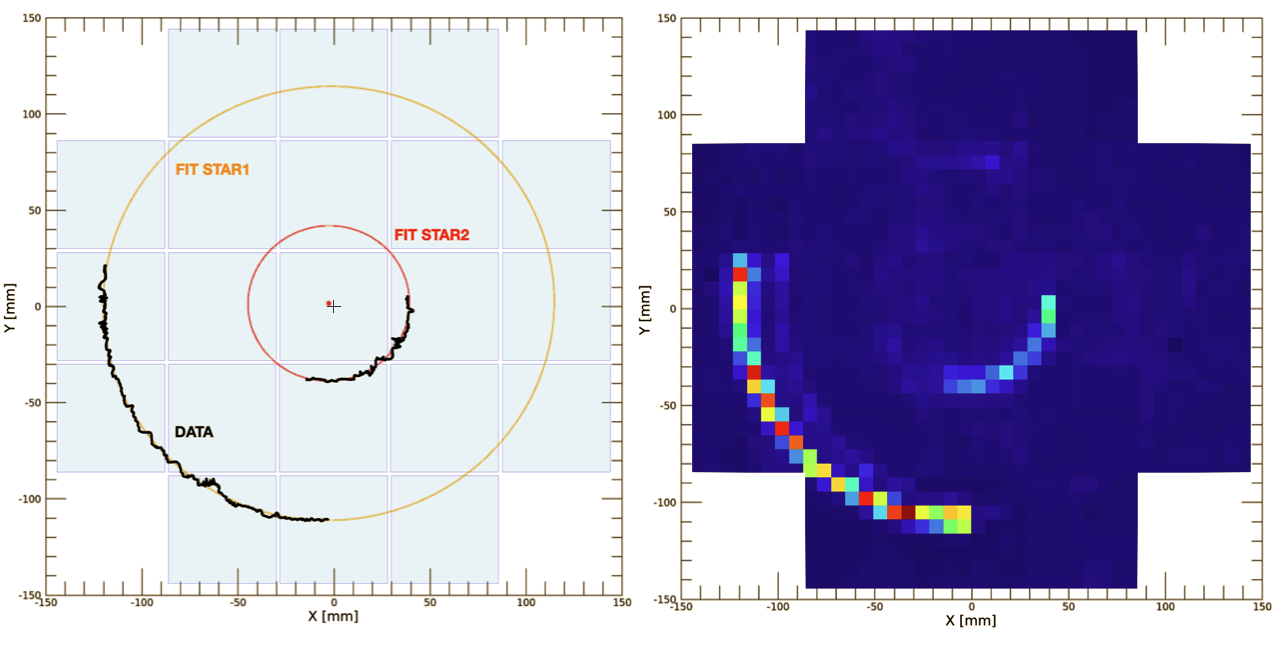}
    \caption{Example of exploitation of the FoV rotation effect. Real data from ASTRI-Horn (right) are fitted with circles (left) to check whether the center of rotation is consistent with the camera geometric center.}
\label{fig_4}
\end{figure}

\noindent
By applying the spherical Laws of Sines to the triangle in figure 2, it can be seen that the value of the parallactic angle $\eta$ is given by the relation \cite{PAR_ANGLE}
\begin{equation}
    \frac{\sin \eta}{\sin\left( \frac{\pi}{2} - \phi \right)} = \frac{\sin h}{\sin z}\,, \label{eq}
\end{equation}
where $\phi$ is the telescope’s latitude, $z$ is the object’s zenith angle (i.e. $\pi/2$ - object’s elevation) and $h$ is the object’s hour angle (local sidereal time - right ascension). As it is clear by relation \ref{eq}, $\eta$ is not a function of time only, but it depends on the pointing direction too. For example, considering a virtual observation of 24 hours, the field of view describes a complete rotation only if the source is circumpolar, i.e. its declination is greater than the latitude of the observer. In all the other cases, the field of view presents a behaviour more similar to an oscillation, as it is shown in figure~\ref{fig_3}. For sure, in no case the rotation of the field of view is uniform.
\noindent
Especially in the context of Cherenkov telescopes, the FoV rotation can be exploited as an effective tool to assess the opto-mechanical behaviour of the instrument. In fact, analysing the trajectories of objects during long exposures, the intrinsic coarse angular resolution can be mitigated using an average process to smooth the artifacts introduced by large pixelization. With this strategy, several characteristics of the camera can be evaluated with high accuracy, using astrometric techniques. For example, the uniformity of the telescope plate-scale can be investigated using long stellar arches at different distances from the center. Furthermore, following the path of a star with constant flux, we can retrieve an efficiency map of the camera considering the fluctuation of the measured signal caused by the crossing of dead areas between the pixels. Finally, another example is the case of the ASTRI-Horn telescope, where the FoV rotation has been exploited to verify post-facto the camera mount system, verifying possible mis-alignments between the camera geometric center and the optical axis of the telescope (center of the apparent rotation), as it is shown in figure~\ref{fig_4} \cite{Iovenitti_effective_pointing}. This approach allowed a direct measurement of a quantity which is very difficult to estimate otherwise and hence this technique has been adopted among the calibration procedures of the incoming ASTRI Mini-Array: dedicated observing runs will be scheduled to exploit the FoV rotation in both the calibration phase and the assembly integration verification \cite{Mineo2021}.

Hence, a need arose for a handy software tool to simulate the effects described above and plan the calibration and assessment campaign. The software presented in this work, “Star Coverage”, is the answer to this need. In the following section, we will present its structure and discuss its output.

\section{Description of the software}
\begin{figure} %
\centering
    \includegraphics[width=0.7\textwidth]{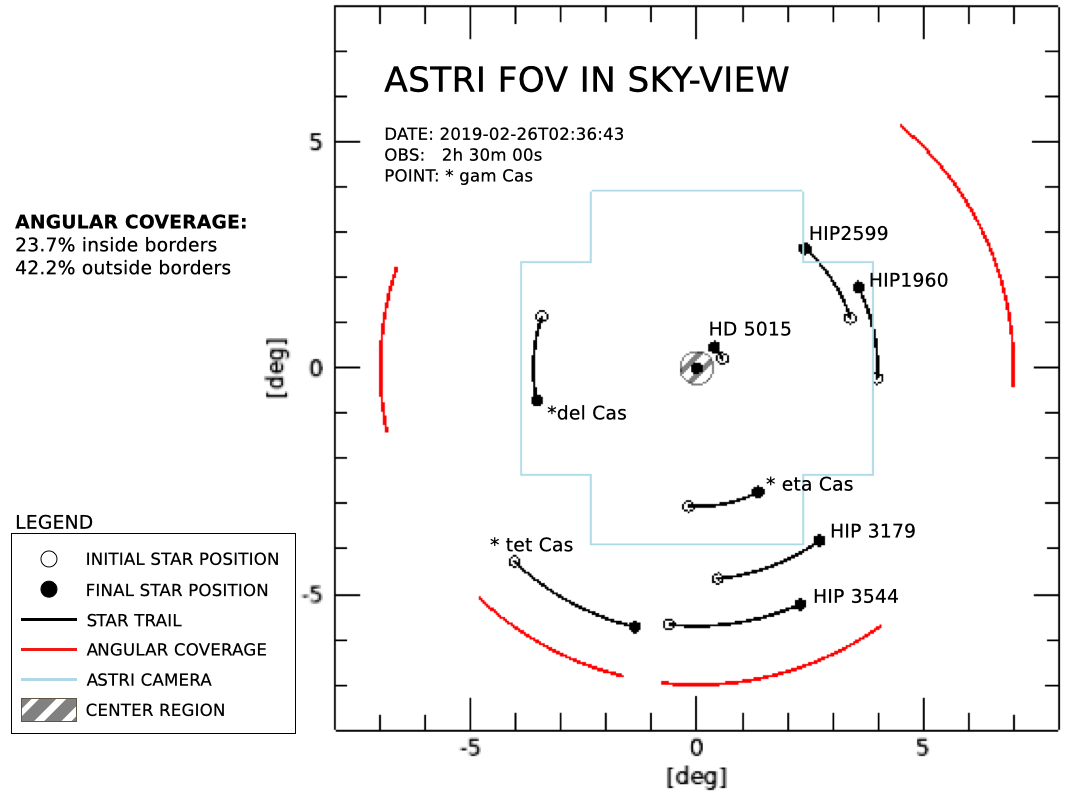}
    \caption{Output of “Star coverage”. The rotation is counterclockwise and the blue polygon is the shape of the ASTRI-Horn camera. Red arches represent the total angular coverage (42.2\%), considering stars outside the borders as well, while the coverage inside the camera would have been 23.7\%.}
\label{fig_5}
\end{figure}

\noindent
“Star coverage” is the name of the handy software we developed to simulate the FoV rotation. It can be used for any telescope, at any time, towards any (reasonable) pointing direction. It is coded in IDL language and the source is available on the web.\\
Given the parameters of the telescope and the observing run, the software simulates the current star field and its rotation in the time interval indicated. In detail, the structure of the program is divided in the steps reported hereafter.
\begin{itemize}
    \item INPUT PARAMETERS. The user must provide: the pointing direction (Ra, Dec) and the radius of the searching area (in degrees); the location of the telescope (lat,lon,el) and its limiting magnitude (see later for the band); the start time, the date and the time length of the observing run. At present there is no graphical user interface, hence the program must be run from the command line, while the parameters must be inserted in the first lines of the source code.
    
    \item QUERY STELLAR CATALOGUES, to get available stars in the searching area, exploiting the Vizier service of the Virtual Observatory\footnote{
        VizieR catalogue access tool, CDS, Strasbourg, France (DOI: 10.26093/cds/vizier).}.
    The default catalogue is Hipparcos (I/239/hip\_main) in the Johnson V band \cite{HIP}.
    
    \item COMPUTE THE (X,Y) POSITION IN THE CAMERA, converting the angular distance into millimeters using the telescope plate-scale\footnote{ 
        Default: ASTRI plate-scale.}.
    
    \item ROTATION OF THE FOV is evaluated solving equation~\ref{eq} every second in the time length indicated by the user.
    
    \item SELECTION OF POINTS. By default, the software considers all the points within the searching area. However, if the geometry of the camera is provided (coordinates of vertices), the points outside the borders are discarded.
    
    \item RESULTS: the output of the procedure is the plot of the FoV rotation. An example is reported in figure~\ref{fig_5}. The percentage of total angular coverage is provided both considering only the points inside the borders and the full searching area.
    
\end{itemize}
\noindent
Optionally, the software can perform additional quality checks on the simulation, if required by the user. In particular, it verifies that the searching area is always above 10° elevation, that the Sun is below -18° elevation (astronomical dawn/dusk) and that the Moon and planets are outside the FoV.

\section{Conclusion}
\noindent
This software, very simple and easy to handle, provides some useful information about the apparent rotation of the stellar component of the night sky background during long tracking observing runs. In particular, it allows the scientists to estimate how many stars below the magnitude limit (in a certain band) fall inside the FoV, the fraction of circle covered by the star rotation and the regions of the camera that will receive the star light.
“Star coverage” constitutes a very general astronomical tool and it can be exploited to simulate the observing runs of any telescope. In particular, regarding the telescopes of the incoming ASTRI Mini-Array, this software will be implemented for scheduling the observing runs both during the calibration and in the assembly verification, but also to optimize the pointing and time length strategy for science observations.

\section*{Acknowledgements}
\noindent
We remember that this work is supported by the Italian Ministry of Education, University, and Research (MIUR) with funds specifically assigned to the Italian National Institute for Astrophysics (INAF) for the development of the ASTRI project. This article has gone through the internal ASTRI review process and we want to thank all the members of that commission. In particular, I would like to thank Giorgia Sironi, my PhD supervisor, and Enrico Giro, for their helpful corrections and constant support.

\end{document}